# POINCARÉ AND RELATIVITY: THE LOGIC OF THE 1905 *PALERMO MEMOIR*

CHRISTIAN BRACCO

UMR Fizeau, Université de Nice-Sophia Antipolis, CNRS, Observatoire de la Côte d'Azur, Campus Valrose, 06108 Nice Cedex, France et Syrte, CNRS, Observatoire de Paris, 61 avenue de l'Observatoire, 75014 PARIS

JEAN-PIERRE PROVOST

Institut Non Linéaire de Nice, Université de Nice-Sophia Antipolis, 1361 Route des lucioles, Sophia Antipolis, 06560 Valbonne, France

We highlight four points which have been ignored or underestimated before and which allow a better understanding of *Sur la dynamique de l'électron*: (i) the use by Poincaré of active Lorentz transformations (boosts); (ii) the necessity, required by mechanics, of a group condition $l=1$ eliminating dilations; (iii) the key role of the action (electromagnetic or not) and of its invariance; (iv) the mathematical status of electron models as example or counter-example.

Following a short communication [1] in June 1905 with the same title, Poincaré develops his approach to relativity in a 59 pages article (an introduction and 9 sections) submitted on July 23$^{rd}$ 1905 [2] and hereafter called the *Memoir*. At first glance, the *Memoir* seems technically difficult and its structure is far from being obvious because of many "flashbacks" expressing Poincaré's hesitations in May 1905. In addition, the electromagnetic models of the electron in §6 nowadays look obsolete and the equation of dynamics (E.D) in §7 seems to depend on them. (It led Miller [3] to consider that Poincaré's relativity is tributary of an electromagnetic vision). Finally, one does not find in [2] Einstein's approach emphasizing the relativity of time, without which there could not be a true understanding of relativity. Poincaré does not rely on the invariance of $c$ (he sets $c=1$), he even does not change his reference frame and amazingly, the contraction of lengths plays a crucial role in his theory because the electron is considered as an extended physical system. We present below four keys, discussed in [4-6], which make the *Memoir* more intelligible and which underline its great merit and originality, namely the central place given to the concepts of group symmetry and action invariance, without which a modern relativistic theory is not conceivable.

In §1, just after having written Maxwell equations and Lorentz force, Poincaré notes that "they admit of a remarkable transformation discovered by Lorentz [invariance of electromagnetism by LT], which is of interest because it explains why no experiment is capable making known the absolute motion of the universe [relativity postulate]":

$$x' = l\gamma(x + \varepsilon t), \quad y' = ly, \quad z' = lz, \quad t' = l\gamma(t + \varepsilon x); \quad \gamma = (1-\varepsilon^2)^{-1/2}. \qquad (1)$$

$l$ is a dilation factor to be specified and $\varepsilon$ is a dimensionless speed (since $c=1$). It is tempting to set $\varepsilon = -v$ to match Eq. (1) with the standard (passive) interpretation of the LT, but this is erroneous since no change of frame is mentioned in [2]. The reason why Poincaré uses active LT (as we have been the first to show it in [4]) is that he has noticed that Lorentz [7] in 1904 could not obtain the full invariance of Maxwell equations because he applied his change of variables to a dielectric boosted in a Galilean way (charges with velocities $v+V$). This explains why, the *Memoir*, which is a mathematical setting of Lorentz' work, begins with a kinematical part (the action of a boost on a moving sphere). Together with the invariance of charge, it allows correcting Lorentz' transformations for electric densities $\rho, j$ and forces $f = \rho(E + v \wedge B)$ and $F = f/\rho$ (those for potentials $A$, $V$ and fields $E$, $B$ being Lorentz' ones). Since $F$ enters the E.D, its law of transformation is a major result of §1. For $\varepsilon = 0$ (pure dilation) it leads to $F' = l^{-2}F$; for $l = 1$, it allows Poincaré suggesting new gravitational forces in §9.

As well known, Poincaré in §4 studies "en passant" the full Lorentz group (dilations included) and its Lie algebra. But his aim being dynamics, as he recalls it, he turns to the trivial demonstration that, if $l$ is a function of $\varepsilon$ and if the transformations still form a group, then $l = 1$. Two questions arise for a modern reader. Firstly, why should $l$ depend on $\varepsilon$? Clearly, because any equation $F = dp/dt$ with $p = mvg(v)$ and $m$ constant (Poincaré sets $m=1$ in §7) cannot be invariant with respect to pure dilations. Secondly, why a group argument? Poincaré's correspondence with Lorentz in Mai 1905 [8] sheds a new light on this question. After having tried $\gamma l = 1$ (conservation of the unit of time) and $\gamma l^3 = 1$ (Langevin's model which satisfies Hamilton equations), he realized that Lorentz' derivation of $l = 1$ (initially obscure to him) dealt with the invariance of the E.D. That such an invariance called for a group property was natural to Poincaré.

The Least action principle (present in §2, 3, 6, 7, 8) and the invariance of action by the LT of Eq. (1) (considered in §3, 6, 8), both play a strategic role in the *Memoir*. As detailed in [5], one must first pay attention to the variables which enter the action because Poincaré makes several reductions of variables, from those of the electrodynamics of fields and charges to those (of position) of a quasi-ponctual electron of Lagrangian $L$, schematically:

$$S = \int d^3r \left[ \frac{(E^2 + B^2)}{2} - j.A \right] dt \xrightarrow{\S 3} \int d^3r \frac{(E^2 - B^2)}{2} dt \xrightarrow{\S 6} \int -L dt. \qquad (2)$$

Secondly, one must take into consideration that the invariance $S = S'$, which is obtained in §3 from the transformations of $E$, $B$ is shown, in a remark of §6, to imply that the Lagrangian $L$ in Eq. (2) must read $L = \gamma^{-1}lL'$. (Primes correspond in §5-6 to an electron set at rest by the boost $\varepsilon = -v$ and $L'$ is a constant). The reason, Poincaré says, is that Eq. (1) implies $dt' = \gamma^{-1}l dt$ if $x'$, or $(x-vt)$, is fixed. If at this stage of the *Memoir*, Poincaré had called for the group condition of §4, he would have obtained immediately (and axiomatically) the relativistic Lagrangian $L = -A\sqrt{1-v^2}$ (written in §7), independently of any model, provided the electron at rest is static.

To understand why Poincaré does not make this axiomatic deduction in the *Memoir*, one must refer once more to his correspondence with Lorentz. In his 2[nd] letter, just having convinced himself that relativity implies in mechanics $l = 1$, he points out a serious problem: Langevin's model, which is in contradiction with the group argument, satisfies $p = \partial L/\partial v$, whereas Lorentz' model with $l = 1$ does not. Therefore a relativistic model of the electron (an existence theorem as mathematicians say) is missing. In a 3$^{rd}$ letter, he announces abruptly that the problem is solved by adding to the electromagnetic Lagrangian *L* a term proportional to the volume of the contracted electron, i.e. to $\sqrt{1-v^2}$. This term ensures the electron stability in a covariant manner. The thorough discussion by Poincaré of models in §6, in relation with the possibility for *l* to be *a priori* different from 1 and with the electron stability, then appears as a pedagogical attempt to let the reader (Lorentz) realize and overcome the problems he himself has encountered. (Lorentz' answer to his reception of the *Memoir* shows that Poincaré has not succeeded).

Once the above four points have been underlined, the derivation in the *Memoir* of the relativistic Lagrangian and the proof of the invariance of the E.D appear to be both general and original. In [5], we compare Poincaré's approach with Planck's later one in 1906. However, is dynamics entirely dealt with in the *Memoir*? Of course it is not. The necessity of a stabilizing term (Poincaré's pressure) will be clarified by von Laue only in 1911 on the basis of relativistic hydrodynamics. More seriously, Poincaré has never thought to include this term in the mass of the electron. The contribution to mass of any internal energy will be understood by Planck and Einstein only in 1907.